\documentclass[a4paper,fleqn]{cas-dc}

\usepackage[authoryear]{natbib}
\usepackage{soul}
\usepackage{amsmath,siunitx}
\def\tsc#1{\csdef{#1}{\textsc{\lowercase{#1}}\xspace}}
\tsc{WGM}
\tsc{QE}
\usepackage{stfloats}
\begin{document}
\let\WriteBookmarks\relax
\def\floatpagepagefraction{1}
\def\textpagefraction{.001}

\shorttitle{Analysis of photometric and spectroscopic variability of Betelgeuse}    

%Kr1: Nechcete pridat doc. Paunzena pripadne doc. Stefla?
%Jad2: přidáno
\shortauthors{D. Jadlovský et al.}  

%Kr1: Ten nazev je prilis vseobecny, na clanek se moc nehodi. Spis treba neco jako "Analysis of photometric and spectroscopic variability of red supergiant Betelgeuse" nebo tak
%Jad2: Změněno
%\title [mode = title]{Photometric and spectroscopic characteristics of red supergiant Betelgeuse}  
\title [mode = title]{Analysis of photometric and spectroscopic variability of red supergiant Betelgeuse}  


%\tnotetext[1]{Photometric and spectroscopic characteristics of red supergiant Betelgeuse} 


\author{Daniel Jadlovský}[type=editor,
       style=chinese,
       orcid=0000-0002-5306-6041] %

%\cormark[1]

%\fnmark[Jadlovský, D.]

%\ead{jadlovsky@mail.muni.cz}

%\ead[url]{<URL>}

%\credit{<Credit authorship details>}

\author{Jiří Krtička}[type=editor,
       style=chinese,]
       
\author{Ernst Paunzen}[type=editor,
       style=chinese,]
       
\author{Vladimír Štefl}[type=editor,
       style=chinese,]

%Kr1:
%\affiliation[]{organization={Masaryk University, Department of Theoretical Physics and %Astrophysics},
%            addressline={Kotlářská}, 
%            city={Brno},
%          citysep={}, % Uncomment if no comma needed between city and postcode
%            postcode={602 00}, 
%            state={South Moravia Region},
\affiliation[]{organization={Department of Theoretical Physics and Astrophysics, Faculty of Science, Masaryk University},
            addressline={Kotl\'a\v rsk\'a 2},
            city={Brno},
            postcode={611 37},
            country={Czech Republic}}

%\author[1,3]{Daniel Jadlovský}[type=editor,
 %      style=chinese,
 %      auid=000,
 %      bioid=1,
 %      prefix=Bc.,
 %      orcid=0000-0000-0000-0000,
  %     facebook=<facebook id>,
 %      twitter=<twitter id>,
 %      linkedin=<linkedin id>,
 %      gplus=<gplus id>]

% Footnote of the second author

% Email id of the second author
%\ead{}



%\affiliation[Masaryk University]{organization={Department of Theoretical %Physics and Astrophysics},
%            addressline={Kotlářská}, 
%            city={Brno},
%          citysep={}, % Uncomment if no comma needed between city and postcode
%            postcode={602 00}, 
%            state={South Moravia Region},
 %           country={CZ}

%\cortext[1]{Jadlovský, D.}



\begin{abstract}
Betelgeuse is a pulsating red supergiant whose brightness is semi periodically variable and in February 2020 reached a historical minimum, the Great Dimming. The aims of this study are to characterize Betelgeuse's variability based on available archival data and to study possible causes of light variability. Many spectra, from ultraviolet and optical regions, were evaluated for spectral analysis. The spectra were used primarily to determine radial velocities from different layers of atmosphere and their long{-}term evolution. Additionally, photometric data were analyzed in different filters as well, to construct light curves and to determine periods of the variability. Spectroscopic and photometric variability are compared to each other and given into a context with the Great Dimming.  

The two most dominant photometric periods are $ P_{1} = 2190 \pm 270 \: \rm d  $ and $ P_{2} = 417 \pm 17 \: \rm d  $, while the dominant optical radial velocity periods are $ P_{1, v_{\rm r}} = 2510 \pm 440 \: \rm d $ and $ P_{2, v_{\rm r}} = 415 \pm 11 \: \rm d $. In the same time, the radial velocity determined from ultraviolet spectra also shows variability and is distinctively different from the variability of photospheric velocity, undergoing longer periods of variability. We attribute these velocities to the velocities at the base of outflowing wind. We also report a maximum of stellar wind velocity during the Great Dimming, accompanied by the previously reported minimum of brightness and the maximum of photospheric radial velocity. After the Dimming, Betelgeuse mode of variability has fundamentally changed and is now instead following a shorter period of $ \sim 200 \: \rm d $.
\end{abstract}

%As the optical (photospheric) radial velocity variations are attributed to the pulsations, this suggests that \hl{both} modes of Betelgeuse's photometric variability are caused primarily by the pulsations
%\item The variability of Betelgeuse during the last 30 years examined in spectroscopy and photometry, using available archival data.
%\item Radial velocity variability in optical and ultraviolet region are distinctively different from each other, while both are partially tied to the brightness variability.
%\item The optical radial velocity variability is tied to both modes of Betelgeuse's photometric variability (about 400 and 2100 days).
%\item In the same time, ultraviolet radial velocity variability is undergoing longer periods of variability. This can be attributed to the velocities at the base of outflowing wind.
%\item Report of maximum of stellar wind velocity in March 2020, during the Great Dimming, accompanied by the historical minimum of brightness and the maximum of optical radial velocity.
%\item \hl{The short and long variability periods} of Betelgeuse's brightness determined, as in accordance with the literature. After the Dimming, Betelgeuse mode of variability has fundamentally changed and is now instead following a shorter period of $ \sim 200 \: \rm d $.

% Keywords

\maketitle

\section{Introduction}
Betelgeuse is a semi{-}variable star and is the brightest star in the near{-}infrared part of spectrum \citep{ori_ir}, typically ranking as one of the 10 brightest stars overall. It is classified as a red supergiant of M1${-}$M2 spectral type \citep{keenan}. However, despite many highest quality observations and research, several of the fundamental characteristics of Betelgeuse remain significantly uncertain.

The uncertainty lies primarily in the determination of the distance of Betelgeuse, and other physical properties tied to it. A reliable estimate of distance determined by \citet{harper_1} $887 \pm 203 \: \rm R_{\odot}$ is based on multi{-}wavelength observations. The analysis by \citet{joyce} that combines evolutionary, asteroseismic, and hydrodynamical simulations, gives a current mass of $16.5{-}19.0\: \rm M_{\odot}$ and initial mass of $18{-}21\: \rm M_{\odot}$.

Betelgeuse has average photospheric radial velocity $ v_{ \mathrm{rad}} = 21.9 \: \rm km \, s ^{-1} $ \citep{famaey}. There have been issues with tracing Betelgeuse back to its birthplace, especially due to the uncertainty of the distance to the star \citep{loon}. Rotational velocity of Betelgeuse is higher than most other red supergiants typically have \citep{loon}. The projected rotational velocity that was determined from HST ultraviolet data of Betelgeuse \citep{uitenbroek} is given as $ v_{ \mathrm{rot}} \mathrm{sin}(i)  \sim 5 \: \rm km \, s ^{-1} $, while a more recent study by \citet{kervella} gives $ v_{ \mathrm{rot}} \mathrm{sin}(i)  = 5.47 \pm 0.25 \: \rm km \, s ^{-1} $. 

Betelgeuse's brightness changes on at least two different timescales \citep{joyce} that were observed since 1837 by sir John Herschel \citep{lloyd}. Based on some of the most precise determinations of the periodicity by \citet{kiss} and \citet{chatys}, the shorter period is $ P_{ \rm short }  \sim 388 \pm 30 \: \rm  days $, and the longer period is $ P_{ \rm long } \sim  5.6 \pm 1.1 \: \rm  years $ ($ \sim $ 2050 days). Both vary in the exact length, and also the amplitude of photometric variability shows significant variability. While there is no clear consensus, it is mostly assumed that the shorter period is driven by atmospheric pulsations in either the fundamental or low{-}overtone modes, and also by oscillations due to invocation of convective cells \citep{kiss}. \citet{joyce} concluded that the mode of atmospheric pulsations is the fundamental mode. The longer period is most often attributed to either flow timescales of giant convection cells \citep{stothers10} or to magnetic activity, rotation of starspots, episodic dust formation, or a nearby companion followed by a dust cloud \citep{wood00,wood04}.

In October 2019 Betelgeuse begun to decrease its brightness once again. However this time, the dimming continued much further than ever before, and it reached the historical minimum by the middle of February \citep{guinan_2}. The unprecedented event was nicknamed as The Great Dimming. The Dimming was first noticed by \citet{guinan}, who suggested that this dimming is due to confluence of the longer and shorter period. Betelgeuse had continued to dim until middle February, reaching a minimum of $ V \sim 1.6 \: \rm mag$. After that it appeared to increase its brightness again \citep{guinan_2}. \citet{gehrz} observed that the Betelgeuse's brightness in infrared was largely unaffected by the Great Dimming, whereas mostly the optical wavelengths were affected. Considering that Betelgeuse is the brightest in the infrared, it suggests that the overall brightness of Betelgeuse remained mostly intact. Therefore it seems unlikely that this episode would be due to major changes within a star, but more likely due to a local surface event.

However, considering that such a dimming had never been observed before, other theories have been proposed besides the conjunction of the two periods. \citet{levesque} argued the dimming episode could not have been caused by a decrease in Betelgeuse's effective temperature, according to model atmospheres. Based on their best fit, the effective temperature dropped only from $3650 \: \rm K$ to $3600 \: \rm K$, which would cause a decrease of visual magnitude $V  \sim 0.17 \: \rm mag$. That would definitely not be sufficient to explain the $V \sim 1.1 \: \rm mag$ drop in brightness. Therefore, they suggest that the decrease in $V$ could be due to mass loss and subsequent large grain dust production that would cause an absorption, mostly in optical part of spectrum. However, \citet{dhamarwa} found out the Betelgeuse has also dimmed in sub{-}millimetre wavelengths by about $20\%$ during the Great Dimming. They argue the Dimming must have been due to changes in the photosphere, as sub{-}millimetre wavelengths are primarily dominated by Betelgeuse's photosphere. Based on their models, they were able to present other possible causes that would explain the scope of the dimming. Either decrease in Betelgeuse's effective temperature to $3450 \: \rm K$, or a significant surface activity through starspots. 

\begin{table*}[htbp]
	\caption{
	List of spectra used for the radial velocity analysis. The dates are given in Modified Julian Date (MJD), which can be obtained by subtracting 2400000.5 days from the Julian date (JD).
	}

	\label{table:spectra} 
	\setlength{\extrarowheight}{3pt}
	\begin{tabular}{ccccc}
\hline

\text{Reference}
& \text{Wavelength region [$\rm \mathring{A}$]} & \text{MJD [d]} & \text{Number of used spectra} &
\text{Source}
\\
\hline
HST GHRS \citep{brandt} & UV (1980{-}3300) & 48889 & 20 & \citet{ghrs} \\
HST STIS \citep{dupree20,ayres} & UV (2270{-}3120) & 50821{-}59476 & 68 &  \citet{stis,astral}\\
HARPS \citep{mayor} & OPT (3780{-}6910) & 58891 & 1 & \citet{eso_arch}\\
UVES \citep{dekker} & OPT (3280{-}5000) & 52530; 56570 & 2 & \citet{eso_arch}\\
X{-}shooter \citep{vernet}  & OPT (2990{-}10190) & 55166 & 1 & \citet{eso_arch} \\
SOPHIE \citep{perruchot}  & OPT (4000{-}6800) & 54074 & 1 & \citet{sophie} \\
ELODIE \citep{baranne}  & OPT (4000{-}6800) & 51204{-}53404 & 8 & \citet{elodie} \\
FLASH HEROS \citep{kaufer}  & OPT (3390{-}8630) & 49868{-}50455 & 4 & \citet{fh} \\
	\hline
	\end{tabular}
\end{table*}

\citet{dupree20} managed to find a connecting link between most of the previous data. Based on the UV observations acquired by HST, they identified a hot structure that had formed in Betelgeuse's southern hemisphere during the beginning of the Dimming. Due to a combination of two major effects, the expansion of photosphere as part of the pulsation period and convective outflows, the apparent mass loss event resulted from an unprecedented powerful outflow. As the material cooled down, the dust opacity could increase quickly enough, which would explain the dimming of Betelgeuse's southern hemisphere. 

The irregularities in Betelgeuse's variability continued, as after its initial rapid rise in brightness that restored Betelgeuse's usual values of brightness, it reached several new minima of brightness, all in shorter time intervals than the previously dominant $ \sim 400 \: \rm  days $ period \citep[AAVSO,][]{dupree22}. A detailed investigation by \citet{montarg21} also supports the mass loss event scenario, although it does not fully rule out the possibility of a decrease in surface temperature. Nonetheless, \citet{harper_2} showed that the Dimming could be explained by a decrease of temperature by about $\sim 200 \: \rm K$, while no additional dust was required.

However, the epoch of Great Dimming and its connection to previous light variability is still not well understood. To better understand the epoch of Great Dimming and the variability of the star in general, we provide here a detailed analysis of photometric and spectroscopic variability covering the last 30 years, i.e., 1990-2021.

\section{Spectral analysis}
\label{chap:spec} 
We analysed numerous high{-}quality calibrated ultraviolet and optical spectra from many publicly available sources using Starlink SPLAT{-}VO \citep{splat}. All the spectra and their sources are listed in Table \ref{table:spectra}. The ultraviolet spectra were acquired by STIS and GHRS spectrographs onboard Hubble Space Telescope (hereafter HST). The optical spectra were acquired by several different spectrographs (see Table~\ref{table:spectra}). In total, about 120 spectral lines were used to determine the radial velocities.

The optical spectra usually include a wide range of wavelengths, therefore it was possible to use one set of spectral lines in SPLAT{-}VO to identify the lines in all the spectra. On the other hand, the ultraviolet spectra were often captured in relatively narrow wavelength windows, therefore it was often not possible to use the same list of spectral lines. Moreover, STIS data are not publicly available in a simple linear format, i.e., a dependency of flux on wavelength. Therefore, in order to analyse the spectra in SPLAT{-}VO, the orders had to be merged. Nonetheless, the spectra that are part of ASTRAL library \citep{ayres} are already converted and publicly available.

HST spectra were captured at different positions relatively to the centre of Betelgeuse \citep{dupree20}, usually scanning from one edge of Betelgeuse to the other edge through its centre. The aperture sizes (given in Table \ref{table:v_r_results}) are smaller than the diameter of Betelgeuse, therefore the position (and size) of the aperture used in each subexposure must be taken into account, as it will affect the results. This does not apply to the optical spectra that we used, because their apertures are larger than the diameter of the star.

Most STIS datasets include 7 subexposures (010{-}070), some up to 9. In the newer datasets the subexposure 040 corresponds to the centre of the star, but in the older datasets other subexposure (or more than one) correspond to the centre (the names of datasets and subexposures are also given in Table \ref{table:v_r_results}). Nonetheless, for the determination of radial velocities we checked the positions of apertures in each dataset to confirm that we used the ones corresponding to the centre of the star. For the analysis of rotational velocity (only for the newer data), we also used subexposures from the ultraviolet surface, usually the edges (010 and 070). The results from these subexposures are given in Table \ref{table:v_rot_results}.

Numerous spectra from HST were omitted based on quality comment of the data, such as due to issues with acquisition of a guide star (the omitted spectra are listed in Table \ref{table:v_r_results}).

\subsection*{STELLA data}
Because the (optical) spectra available to us contain large time gaps, we also adapted high{-}quality determinations of radial velocities by STELLA robotic spectrograph \citep{granzer21} to improve the results, as they aggregated considerably more data. This dataset consists of unprecedentedly enormous amount of the radical velocity determinations in optical region, about 2000 spectra across 14 years (covering MJD = $54754{-}59716$) . The instrument and data reduction are described in \citet{stella,stella_2}.STELLA data are used in Chapter \ref{chap:rl_curve}.  

\subsection{Radial velocity}
\label{chap:v_rad} 

Spectral lines used for the radial velocity analysis were selected primarily based on \citet{carp}, \citet{brandt} and \citet{freeshell}. For determination of radial velocity, we used only such lines that could be easily identified even in spectra of lower quality. The spectral lines that we used are given in Table \ref{table:lines}. To determine radial velocity, we primarily used selected absorption lines and centrally reversed emission lines (in UV). Afterwards, the lines were fitted with SPLAT{-}VO, using Voigt profile. 

As the ultraviolet spectra consist of various types of spectral lines, we divided ultraviolet spectral lines into two groups \citep{brandt,carp} that were analysed separately. The first group is composed of lines that are supposed to originate in the warm chromosphere or at the base of outflowing wind as given in paper by \citet{carp}. The lines in this list span from far ultraviolet to $ 2881 \: \rm \mathring{A}$. As the majority of ultraviolet spectra in our sample start at about $2300 \: \rm \mathring{A}$, we did not use all the lines. Some weaker lines were also not used. Almost all the lines in this list correspond to single ionized metals and a large portion of them also have emission wings (centrally reversed emission). To determine radial velocity, we analysed the central absorption feature, therefore these results correspond to velocity of stellar wind \citep{wood04, wood16}, and thus we will refer to them as such. Analysis of other absorption features in this region gave similar results of radial velocity, but in further analysis we used only the lines from \citet{carp}.

The second group of lines in the ultraviolet spectra are from a wavelength region above $2900 \: \rm \mathring{A}$, where the photospheric absorption starts to dominate the spectrum and photospheric continuum becomes prominent (although some chromospheric features remain significant) \citep{brandt,carp}. Spectral lines in this group consist primarily of absorption by neutral metals. The radial velocity determined from this group should correspond to outer atmosphere \citep{brandt}.We shall refer to this group of lines as ultraviolet photospheric region.

To test the validity of this approach, we also divided the lines into groups based on their excitation potential (in the UV and optical region) and fitted them separately, but the differences between determined radial velocities did not exceed the margin of error. For the stellar wind lines we also compared the absorption cores velocities to velocities of the emission wings centroids, but it gave similar results. Lastly, we determined radial velocity from pure emission features (without central reversals). These determined velocities were higher than the ones from stellar wind lines, usually centered at rest with respect to the photosphere, which is in agreement with \citet{carp_2}. However, we were able to analyse only a few of such emission lines therefore we will not use them in the following analysis.
%(including the Balmer lines)
%Likewise, when the results in ultraviolet spectra are grouped based on whether they come from absorption or centrally reversed emission lines, there does not appear to be a systematic difference between them. 

\begin{figure}[htbp]
    \centering
    \includegraphics[width=0.5\textwidth]{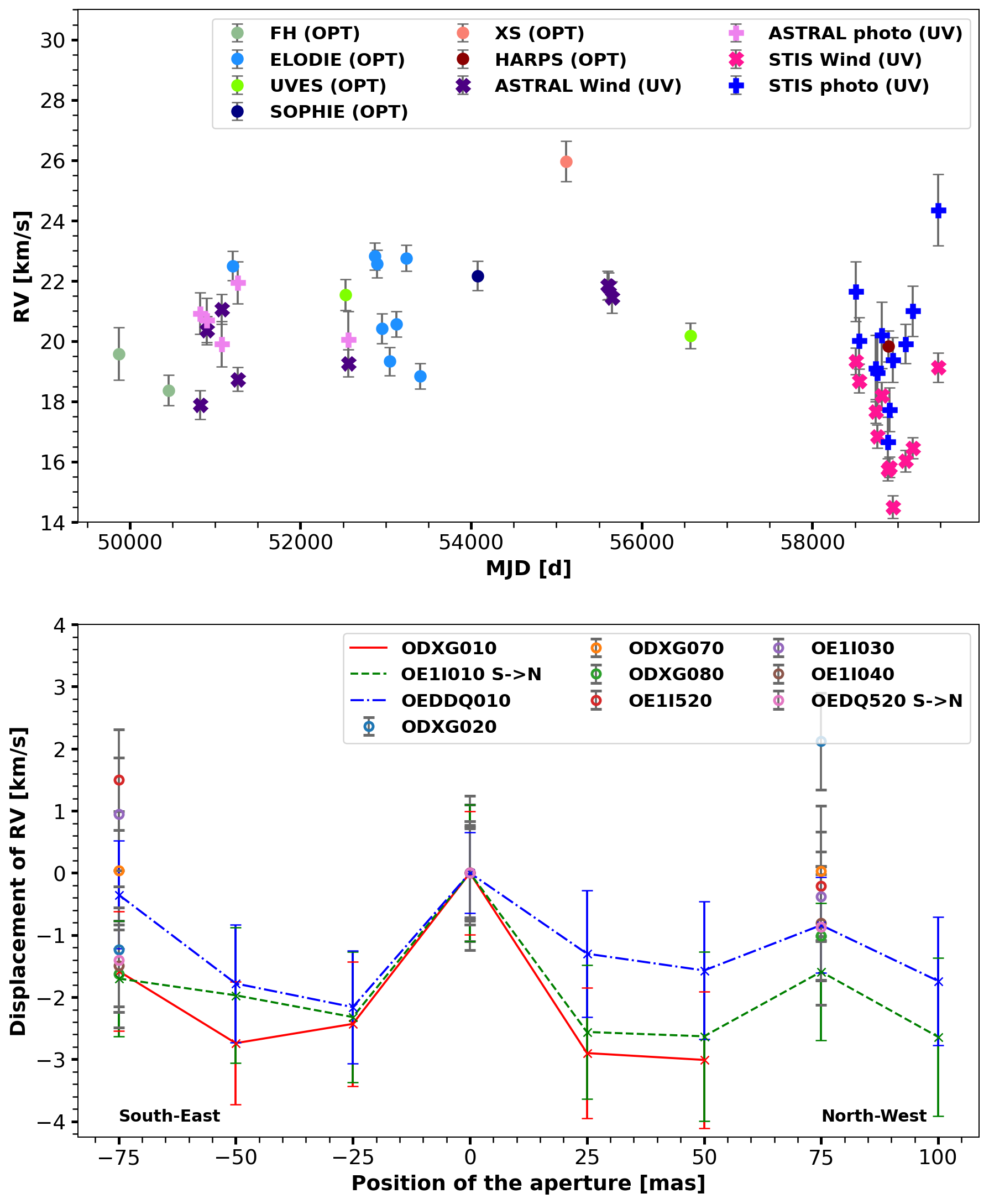}
    \caption{
    {\em Upper panel:}
    Dependence of radial velocity $ v_{ \rm r} $ on MJD, using results from all the spectra that were analysed in this paper (Table \ref{table:v_r_results}). 
    {\em Bottom panel:}
    Radial velocity from the ultraviolet photospheric region measured across the disk of Betelgeuse from the datasets \textbf{ODXG}, \textbf{OE1I} and \textbf{OEDQ}, as given in Table \ref{table:v_rot_results}. The position is relative to the disk center and position angles of scans were not the same each time. Position angles of the datasets labeled as S->N were orientated about $\sim 30  ^{\circ}$ from the position angle of the axis of rotation, therefore the observed rotational effects should be minimal. The zero value on y coordinate corresponds to the central subexposure (Table \ref{table:v_r_results}).
    %{\em Bottom panel:}
    %Optical radial velocity $ v_{\rm r} $ phased by period $ P_{1, v_{\rm r}}  = 383 \pm 7 \: \rm d $. The symbols with error bars correspond to the optical radial velocities determined here, while the symbols without error bars are the ones adapted from \citet{granzer}. 
    }
    \label{fig:vr_vrot}
\end{figure}

The determined radial velocities are listed in Table \ref{table:v_r_results} and are plotted together in Figure \ref{fig:vr_vrot}. The HST STIS datasets \textbf{ODXG}, \textbf{OE1I} and \textbf{OEDQ} are the ones near the Great Dimming. We also used them to analyse rotational velocity by using the subexposures across the surface of the star. The individual radial velocities for each of the subexposures are listed in Table \ref{table:v_rot_results}. In ASTRAL \textbf{O4DE} dataset, i.e., the one between MJD 50821{-}51265, we only used subexposures 050 as they correspond to the centre of the star. On the other hand, subexposures in the ASTRAL dataset \textbf{OBKK} use different wavelength regions rather than positions on the surface, as well as slightly larger apertures. Most of the spectra within this dataset that we used are within the wavelength region under $ 2900 \: \rm  \mathring{A} $,  but do not cover the photospheric region.

In further analysis, some UV spectra were omitted due to significantly larger apertures compared to other UV spectra, such as spectra from GHRS and OBKK74010, as the different aperture size might affect the resulting radial velocity (and it indeed appears to be the case). \textbf{O6LX} and \textbf{OBKK} datasets also have larger apertures, but less than the omitted spectra, and it appears the determined radial velocities from this dataset are of similar values as most of other datasets that we used, although they appear slightly larger.

The data plotted in Figure \ref{fig:vr_vrot} have some considerable gaps, but fortunately they seem to cover the full range of possible radial velocities $ v_{ \rm r} $ values, i.e., the minima and maxima. At several places, the results also display a variability at smaller time scales, such as the ones from ELODIE, and especially the ones from STIS. 

The radial curve shows a major systematic difference between the velocities determined from the optical and ultraviolet spectra. While the optical radial velocities $ v_{ \rm r} $ have a range between about $19 {-} 28 \: \rm km \, s ^{-1} $, the velocities of stellar wind have a range of only $14 {-} 22 \: \rm km \, s ^{-1} $. The variability itself appears to be happening on a different time scale as well. For example a closer examination of the well covered ultraviolet peak (in both ultraviolet line lists) of ASTRAL \textbf{O4DE} dataset (the values with MJD between 50821{-}51266) shows a relatively slow variability near the local radial velocity maximum, covering roughly 370 days, while the profile of the peak suggests that the period is much longer. On the other hand, on a similar timescale radial velocities from ELODIE appears to undergo one full period of variability. Similarly, examining the \textbf{ODXG}, \textbf{OE1I} and \textbf{OEDQ} ultraviolet radial velocities near The Great Dimming, covering about 820 days, yields a slow fall into a minimum, culminating by the time of the Great Dimming, and a subsequent slow rise. Therefore, the data suggest that radial velocity in ultraviolet is undergoing a variability on a much longer time scale than radial velocity in the the optical part of the spectrum, and on top of that the amplitude is not the same. The photospheric ultraviolet radial velocity ($ 2900 {-} 3100 \: \rm  \mathring{A} $) is more similar to the optical radial velocity, at least in terms of amplitude. Considering the event that is covered by \textbf{O4DE} dataset, when the the radial velocities from the two ultraviolet regions intertwined, it shows that both groups of ultraviolet lines that we selected are indeed fundamentally different and have different periods of variability.

In optical region, the changes of the radial velocity  we observe are largely due to the pulsations of Betelgeuse. Henceforth, the variability is due to some layers of Betelgeuse's atmosphere changing its relative velocity towards us. As the majority of contributions to the spectra are from the photosphere, it would be more relevant to use term photospheric velocity to describe the velocity variability.

\subsection{Rotational velocity}
As discussed previously, the STIS spectra are taken from different parts of Betelgeuse surface including the centre and edges. Thus, it could be possible to unveil rotational effects by comparing the spectra from different parts of Betelgeuse surface as one side of Betelgeuse should be moving towards us, while the other side away from us. Table \ref{table:v_rot_results} lists ultraviolet radial velocities determined from different parts of the stellar surface, which are plotted in the bottom panel of Figure \ref{fig:vr_vrot}.

%We plotted only results from the ultraviolet wavelength region $ 2800 {-} 3100 \: \rm  \mathring{A} $, as the other ultraviolet region did not seem to have a systematic trend.

While the the resulting plots (bottom panel of Figure \ref{fig:vr_vrot}) show a systematic trend, it does not reveal the projected rotational velocity $ v_{\rm rot} \sin(i) $, as both edges of Betelgeuse seem to be moving towards us mostly at a similar velocity. According to \citet{kervella}, the value of projected rotational velocity is $ v_{ \mathrm{rot}} \mathrm{sin}(i)  = 5.47 \pm 0.25 \: \rm km \, s ^{-1} $, therefore the effects should have been theoretically observable. But based on the data, it is questionable whether rotational effects are observable in this region of UV. That notion is reinforced by the fact that ultraviolet diameter of Betelgeuse is much larger than the optical diameter \citep{dupree20}, therefore distances of most subexposures are further from the centre than the diameter of photosphere.

\section{Photometry}

\begin{table*}[htbp]
	\caption{List of photometric data used for the light curve analysis.}
	\label{table:photometry} 
	\setlength{\extrarowheight}{3pt}
	\begin{tabular}{cccc}
\hline
\text{Reference} & \text{Wavelength region [$\rm \mathring{A}$]} & \text{MJD [d]} & \text{Source}

\\
\hline
AAVSO \citep{aavso} & 3000{-}19370 & 48500{-}59800 & \citet{aavso} \\
BRITE \citep{pig} & 3900{-}4600 \& 5500{-}7000 & 56500{-}59250 &  \citet{brite}\\
SMEI \citep{jackson} & 4500{-}11000 & 52500{-}55800 & \citet{smei_data} \\
	\hline
	\end{tabular}
\end{table*}

For photometric analysis, the data from AAVSO, BRITE and SMEI were used (Table \ref{table:photometry}).

\subsection{Light curve}
\begin{figure*}[htbp]
    \centering
    \includegraphics[width=1.0\textwidth]{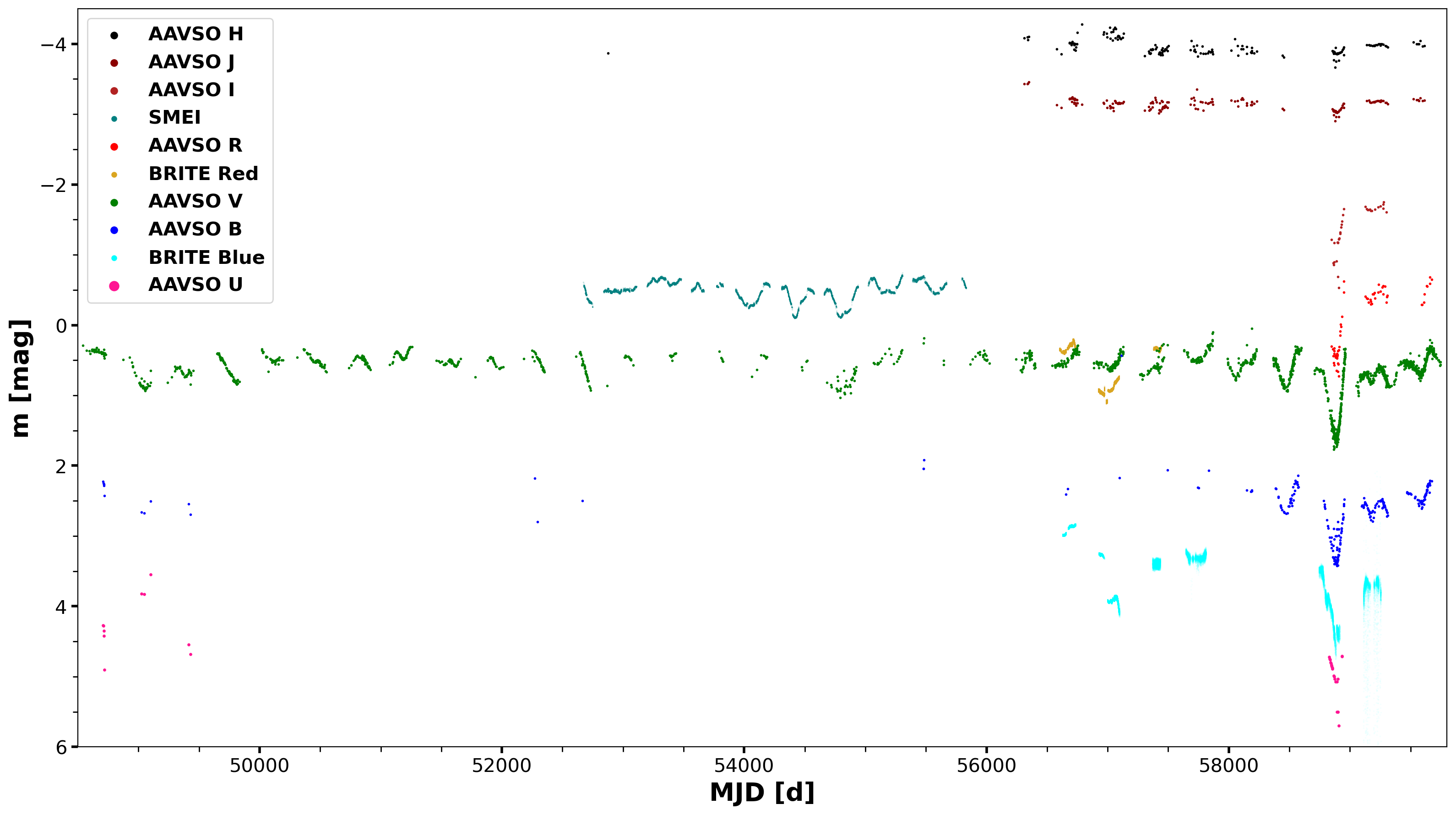}
    \caption{Light curve comprising of all the photometric data adopted here, from H (upper curve) to U (bottom curve), i.e., in the same order as in the legend. The data cover the last three decades and end shortly after The Great Dimming. The data from $ V $ filter range even farther into history, but they were not plotted to improve scaling.  }
    \label{fig:photo_all}
\end{figure*}

Combining the data gives us almost a continuous brightness variability of Betelgeuse for the last two decades, with SMEI covering the first decade, and AAVSO supplementing sufficiently enough the rest. The combined plot can be seen in Figure \ref{fig:photo_all}.

As can be seen in Figure~\ref{fig:photo_all}, the observations in filters using longer wavelengths ($I, J, H$) are considerably less affected by the variability of the star \citep{gehrz}. As the star is the brightest in near{-}infrared region, this means that the variability we observe (most often in optical region) does not affect the overall brightness of the star in such a dramatic way. That can be seen in even greater extent during the Great Dimming as well, which suggests several possible causes of the Dimming. For the determination of period, SMEI is the best source, as it has a continuous high{-}quality coverage of almost two periods. The $ V $ filter from AAVSO is also very promising, but it has large time gaps. The time gaps are caused by the position of the star, as during summer period in the northern hemisphere it is difficult to observe Betelgeuse by ground based observations (therefore most of the optical spectra have been 
acquired in winter as well).

\subsection*{SMEI data}
The Solar Mass Ejection Imager \citep[SMEI,][]{eyles} was launched into an Earth-terminator, Sun{-}synchronous, $840 \: \rm km$ polar orbit as a secondary payload on board the Coriolis spacecraft in January 2003 and was terminated in September 2011. Its main purpose was to monitor and predict space weather in the inner solar system. SMEI comprised three wide{-}field cameras, which were aligned such that the total field of view is a $180 \: \rm deg$ and about $3  \: \rm  deg$ wide arc, yielding a near-complete image of the sky after about every $101.5  \: \rm  min$  orbit. 

A detailed description of the data analysis pipeline used to extract light curves from the raw data is provided by \citet{hick}. Because of the presence of strong instrumental trends, a basic cleaning algorithm as described in \citet{paunzen2021} was applied.

\section{Period Analysis}
\label{chap:rl_curve}
The following periods were determined using CLEANest method in the PERANSO program \citep{peranso}. It features several methods to deal specifically with irregularly spaced data and multi-period signals (it was also used to clean the BRITE and SMEI data by removing outliers). As we can see in Figures \ref{fig:radial_light} and \ref{fig:rl_dimming}, the variability has fundamentally changed during the Dimming, therefore in the following analysis the period is determined separately before and after the Great Dimming.

\subsection{Photometric variability}
Photometric period analysis was performed using the SMEI and AAVSO V data in PERANSO, after outliers were removed (for SMEI). The dominant periods (before the Dimming) are $ P_{1} = 2190 \pm 270 \: \rm d  $ and $ P_{2} = 417 \pm 17 \: \rm d  $, $ P_{3} = 365 \pm 75 \: \rm d  $ and $ P_{4} = 185 \pm 4\: \rm d  $. The third strongest period $ P_{3} $ shares the peak with $ P_{2}$. These results correspond well to the values given by \citet{kiss} and \citet{chatys}, except $ P_{4}$, which might correspond to observational gaps in summer. The longer period is often given with a similarly large uncertainty, such as in the literature cited above. After the Dimming, the dominant mean period is $ 230 \pm 29 \: \rm d $.
% However, as SMEI covers $ \sim $ 3000 days, it is questionable whether PERANSO's determination of $ P_{1}$ is reliable.
\subsection{Radial velocity variability} 
As was discussed in \autoref{chap:v_rad}, radial velocity from different parts of Betelgeuse's spectrum, i.e., ultraviolet and optical, should be examined separately.

The dominant periods of optical region (before the Dimming) are $ P_{1, v_{\rm r}} = 2510 \pm 440 \: \rm d $, $ P_{2, v_{\rm r}} = 415 \pm 11 \: \rm d $,  $ P_{3, v_{\rm r}} = 185 \pm 4 \: \rm d $ and $ P_{4, v_{\rm r}}  = 370 \pm 9 \: \rm d $\footnote{Even without using the additional data from STELLA it was possible to detect the shorter period by \citet{kiss} and \citet{chatys}}. The second and fourth results correspond well to the shorter period given by \citet{kiss} and \citet{chatys}, as well as to photometric periods $P_2$ and $P_3$ determined here (they share the same peak as well). The period $ P_{1, v_{\rm r}} $ shows a clear connection to the longer period. After the Dimming the dominant mean period is $ 200 \pm 18 \: \rm d $, which is perhaps connected to the period $ P_{3, v_{\rm r}}$ that was present before the Dimming and is the same (within margin of error) as the post Dimming period determined from the photometry.

%Therefore, considering that the variability of radial velocity, i.e., the photospheric velocity, is attributed to the pulsations of the atmosphere, these results support that \hl{both main modes of the variability are connected to} Betelgeuse's pulsations.
%\hl{The radial velocities determined from the optical region are phased by} $ P_{2, v_{\rm r}}$ \hl{in the bottom panel of} Figure \ref{fig:vr_vrot}.
%But in general, the variability of optical radial velocity does not fully correspond to the longer period, as it is only the fourth strongest period... while the longer period is due to other processes that may partially affect radial velocities as well. 

The dominant periods determined from the velocity of stellar wind are: $ 1432 \pm 76  \: \rm d $, $ 2410 \pm 470 \: \rm d $, and $ 757 \pm 16  \: \rm d $. The second period suggests a connection to the longer period. The other two peaks are relatively strong and suggest different new periods. Nonetheless, there are no significant peaks similar to the shorter period, therefore there does not seem to be a connection of a chromosphere or stellar wind to the shorter period.

For the ultraviolet photospheric region the determined periods are: $ 560 \pm 12  \: \rm d $, $ 1190 \pm 50 \: \rm d $, and $  653 \pm 17  \: \rm d $. The first period is to a degree close to the periods determined from optical radial velocities $ P_{2, v_{\rm r}}$ and $ P_{4, v_{\rm r}}$, which suggests that in this wavelength region the photosphere does indeed become prominent. The other periods are somewhat similar to the shorter periods of stellar wind velocities, although there are numerous narrow peaks. 

Therefore, it appears that the variations of ultraviolet radial velocities are not caused by the pulsations, or at the very least not by the same processes that drive the variability of radial velocity in the optical region. Instead, the radial velocity determined from the ultraviolet photospheric region is likely a subject to other processes in outer parts of the Betelgeuse's atmosphere, possibly loosely connected to the processes that lead to the longer period, while the main group of the ultraviolet lines represent the stellar wind (as discussed previously). Furthermore, \citet{kervella} showed that the angular velocity changes as a function of radius, so the results are likely affected by that as well.

\subsection{Connection of radial velocity and light variations}
\begin{figure*}[htbp]
    \includegraphics[width=0.75\textheight, height=1.0\textwidth, keepaspectratio]{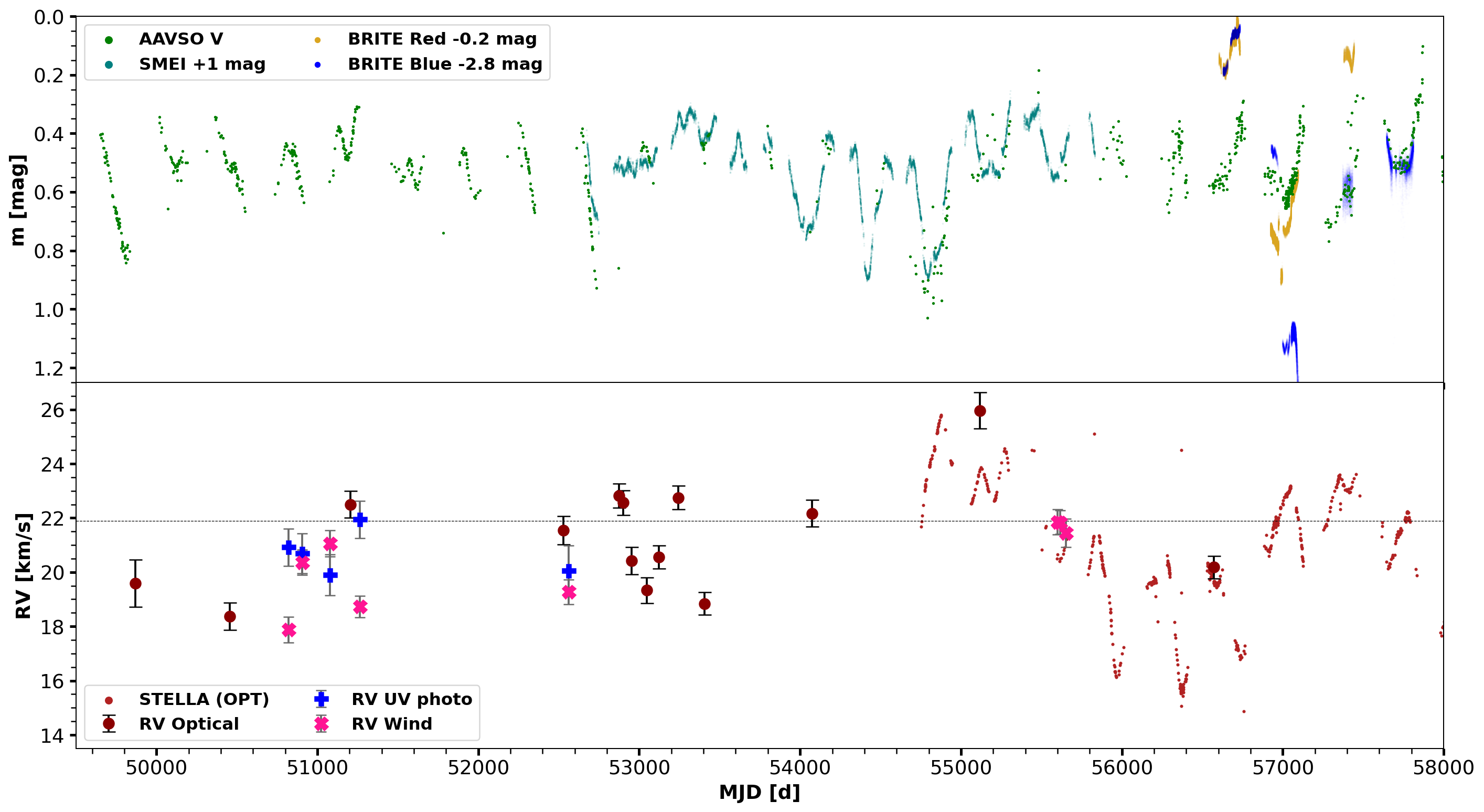}
    \caption{
    Joint plot of radial  velocities and photometric data covering the last 30 years, except the Great Dimming. The plot of photometric data ({\em upper panel}) contains only data in selected filters due to scaling issues. The plot of radial velocities ({\em lower panel}) includes the data from \citet{granzer21}. RV Optical refers to the optical spectra that were analysed in this paper. The broken dashed line corresponds to Betelgeuse's rest velocity $ 21.9 \: \rm km \, s ^{-1} $ \citep{famaey}. The plot is continued in Figure \ref{fig:rl_dimming}. }

    \label{fig:radial_light}
\end{figure*}

\begin{figure}[htbp]
    \centering
    \includegraphics[width=0.5\textwidth]{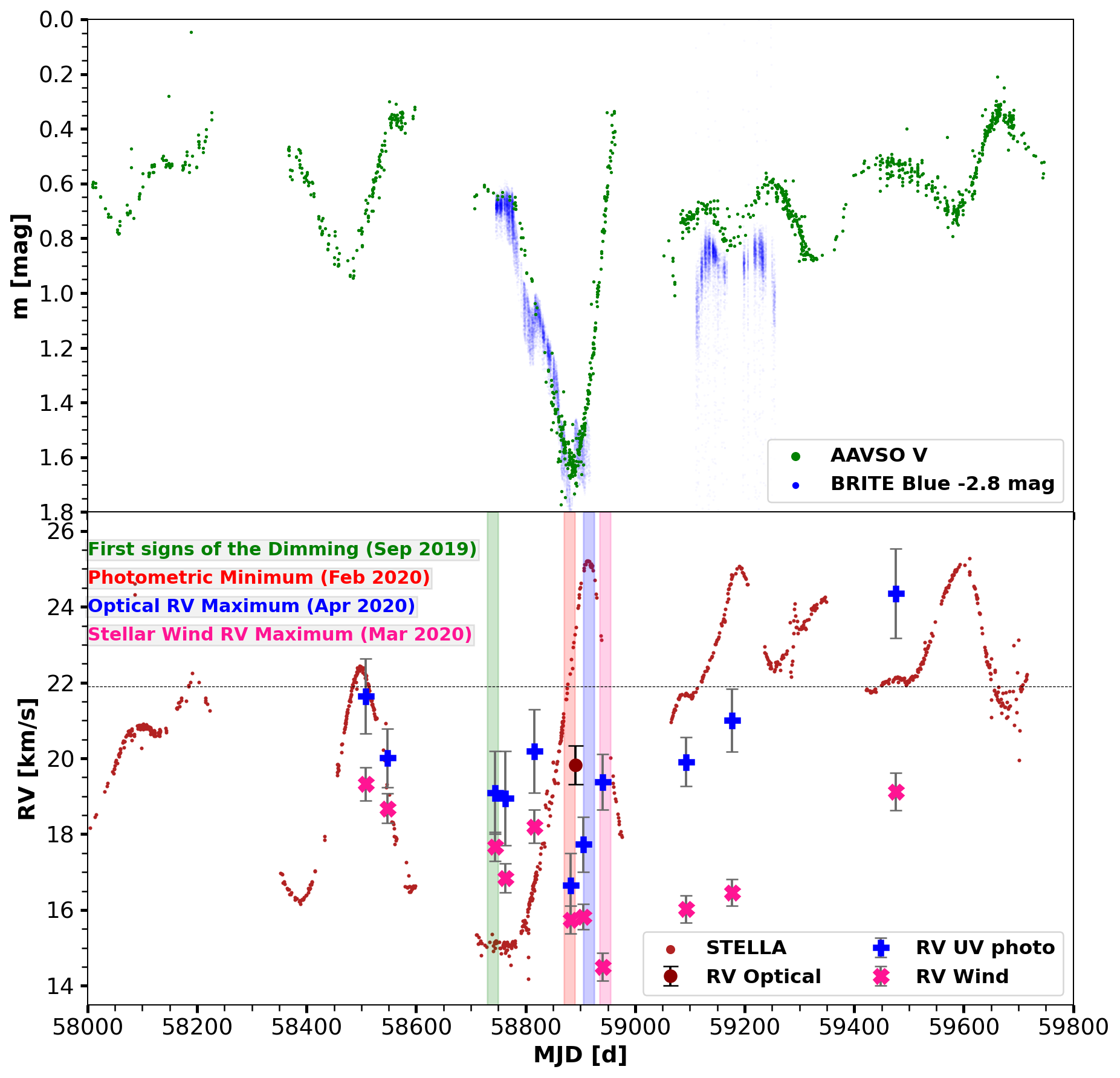}
    \caption{
    Detailed plot of magnitude variations ({\em upper plot}) and radial velocity ({\em lower plot}) during the period of The Great Dimming and after the Dimming. Plotted using additional data by \citet{granzer21} and marking the significant points in time.
    }
    \label{fig:rl_dimming}
\end{figure}

The radial velocity and major photometric data were plotted together in Figure \ref{fig:radial_light} and \ref{fig:rl_dimming}, so that the curves can be directly compared. Out of the optical radial velocities analysed here, only 3 were taken during the same period as STELLA. Considering that different methods were used, our results correspond reasonably well to the adapted values adopted from from literature, although the velocity determined from X{-}shooter seems to be too high. The variability of amplitudes of both curves is clearly correlated, i.e., when the amplitude of magnitude is smaller, the amplitude of radial velocity is usually also smaller and it is apparent that both curves follow similar periods of variability.

Furthermore, the maxima of optical radial velocity seem to be usually delayed by roughly one month after the brightness minima. As we can see in Figure \ref{fig:rl_dimming}, during the Great Dimming the velocity of stellar wind reached a maximum. The ultraviolet radial velocity from the photospheric region is indeed more similar to the optical radial velocity, sometimes it appears to follow the optical light curve very closely, such as during the beginning of the Great Dimming. Nevertheless, during other parts of the Great Dimming it is clearly not following the optical radial curve anymore. 

\section{Discussion and conclusions}
In this research we aimed to study Betelgeuse’s variability in pursuit of explaining its causes, and possibly to bring some context to the unprecedented Great Dimming. To accomplish that, we accessed many available public archival data, coming from various high-quality instruments. The results show that variability of optical and ultraviolet radial velocities are distinctively different from each other, as well as the brightness variability in various filters. These results and their comparison allow us to study the variability in a greater extent.

The optical radial velocity variability corresponds very well to both periods given by \citet{kiss} and \citet{chatys}. The changes of radial velocity are likely caused by the pulsations of the photosphere, thus we interpret the radial velocity from optical region as a photospheric velocity of Betelgeuse. The longer mode of variability is also prominent in the optical radial velocity, therefore it appears that convection strongly influences the radial velocity as well, as the longer period is often attributed to convection cells \citep{stothers10}.
%Such a period corresponds to the values determined by \citet{kiss} and \citet{chatys}. 

In the ultraviolet spectra we analysed two groups of spectral lines separately, stellar wind lines and lines from a region where the influence of the photosphere rises \citep{carp}. Despite the fact that most of the spectra studied here were obtained in the ultraviolet region, the data was usually restricted to relatively short periods of time, thus proving it difficult to correctly determine the overall period. 

The systematic difference between the values of optical radial velocity and the stellar wind lines is primarily caused by the expansion of stellar wind. As the stellar wind accelerates material towards us, the radial velocity determined in upper parts of atmosphere is therefore lower, relative to Betelgeuse's rest velocity. Our results for the velocity of stellar wind are similar to the ones reported by \citet{carp_2, carp}, who report velocity outflows of several $ \rm km \, s ^{-1} $, increasing with optical depth of a given line. However, in their studies they analysed shorter periods of time. As we analysed larger sample of data, we show that the velocity of stellar wind is highly variable as well, at some times it is close to the star's rest velocity. The variability of stellar wind is likely affected by a multitude of processes in outer parts of atmosphere. It likely corresponds to various episodic gaseous outflows \citep{humph}, local (up/down)flows and global nonradial chromospheric oscillations that might partially be in phase with photospheric pulsations \citep{lobel}. Nonetheless, the time scale of the variability is different than for the photospheric radial velocity. We analysed periodicity of stellar wind and found that periods $ 1432 \pm 76  \: \rm d $ and $ 2410 \pm 470 \: \rm d $ are the most prominent, while the second one is similar to the longer period derived by \citet{kiss} and \citet{chatys}.

The ultraviolet photospheric region did not yield a reasonably accurate period, although the periods and amplitudes are considerably closer to the periods of photospheric radial velocity. To a certain degree, the radial velocity appears to be affected by the overall magnitude variability and pulsation cycle. The possible explanations for the processes that drive the variability of these groups of lines are less obvious. Partially, it can be affected by the same multitude processes as discussed in the previous paragraph. Possibly, it could also correspond to velocity of stellar wind, but closer to the photosphere, where the speed of stellar wind is lower. Besides that, this region is undoubtedly affected by the photosphere significantly, as at some occasions its radial velocity is following the optical radial curve and is of similar values. However, the results of velocity based on the ultraviolet spectra are not as abundant and precise as the measurements from STELLA, therefore the interpretation of these results remains rather ambiguous.

As part of studying the ultraviolet spectra, we also attempted to measure projected rotational velocity $ v_{\rm rot} \sin(i) $ of Betelgeuse, by studying additional spectra from the disk of the star, and comparing them to central spectra. We used the results from the ultraviolet photospheric region, where the photosphere is already prominent. But this analysis did not provide reasonable results. Apart from other reasons already discussed, the reason that the method did not succeed might be largely due to the fact that the parts of Betelgeuse's atmosphere where ultraviolet spectral lines that we used are formed simply do not fully reflect the rotation of photosphere. However, \citet{kervella} suggests a period of rotation $ 31 \pm 8$ years and a rotational coupling of Betelgeuse and its chromosphere, therefore if a more elaborate method would have been used, it would have likely been possible to determine $ v_{\rm rot} \sin(i) $. Rather than measuring rotational effects, our results of radial velocity across the Betelgeuse's disk probably corresponds to the local upflows and downflows with amplitudes up to $ \sim 2 \: \rm km \, s ^{-1} $ reported by \citet{lobel}. This suggests why the results of ultraviolet radial velocity in this study did not provide expected results of $ v_{\rm rot} \sin(i) $.

As part of photometric analysis, it was possible to reliably determine the periods $P_{1} = 2190 \pm 270 \: \rm d  $ and $ P_{2} = 417 \pm 17 \: \rm d $, which is in high accordance with the most dominant periods determined by optical photospheric velocities. These photometric periods are also in accordance with the values by \citet{kiss} and \citet{chatys}. The photometric data do not support that the Great Dimming is due to a simple convergence of the shorter and longer period, as the amplitudes of the two main periods could not cause such a decrease in brightness even when combined. Fortunately, the AAVSO photometric data from various filters (Figure \ref{fig:photo_all}) give a major information about what could have possibly caused the Dimming. Based on the filters from near{-}infrared region, where Betelgeuse is the brightest, the Dimming is on a much smaller magnitude scale (filter $ I $) than in optical filters, or barely noticeable at all (filters $ J $ and $ H $) \citep{gehrz}. Therefore, this possibly means that Betelgeuse was actually not physically affected by the Great Dimming, i.e., that the star did not change its luminosity, but that there was an extinction, mostly in optical region. That could have been due to dust particles in a proximity of the star, therefore Betelgeuse must have experienced a significant loss of mass prior to the Dimming. The reason could also be that Betelgeuse's effective temperature $ T_{ \rm eff} $ decreased. That would cause the peak of Betelgeuse's brightness to move to longer wavelengths, henceforth the Dimming would be of a smaller scale in the infrared. So Betelgeuse could have decreased either its entire surface's effective temperature, or only some parts of the surface did. Betelgeuse is known to have large convective cells and starspots \citep{kervella}, so possibly the overall surface temperature could have temporarily decreased due to an unprecedented activity of this kind.

Most importantly, the spectral analysis reveals the global minimum of ultraviolet radial velocity (in both UV groups) during the Great Dimming, therefore the velocity of stellar wind had its maximum. Either way, this supports that during the Great Dimming there was a massive outflow of material, most likely connected to the mass loss event in southern hemisphere of Betelgeuse that was reported by \citet{dupree20,kravchenko21,montarg21} and regarded to be responsible for the Great Dimming. It was followed by the minimum of brightness in February 2020 \citep{guinan_2} and later by a maximum of optical radial velocity in April 2020, as reported by \citet{granzer21}. Furthermore, in Figure \ref{fig:rl_dimming} we can also see that after the Great Dimming Betelgeuse did not return to its original mode of variability, but it is now instead following a much shorter period of $ \sim 200 \: \rm d $. These findings give us new insights regarding The Great Dimming, the connection between Betelgeuse's photosphere and chromosphere and of nature of stellar wind in red supergiants that should be examined further.

\section*{Acknowledgements}
Based on data collected by the BRITE Constellation satellite mission, designed, built, launched, operated and supported by the Austrian Research Promotion Agency (FFG), the University of Vienna, the Technical University of Graz, the University of Innsbruck, the Canadian Space Agency (CSA), the University of Toronto Institute for Aerospace Studies (UTIAS), the Foundation for Polish Science and Technology (FNiTP MNiSW), and National Science Centre (NCN).

Based (partly) on data obtained with the STELLA robotic telescopes in Tenerife, an AIP facility jointly operated by AIP and IAC. The authors are grateful for the permission to use yet unpublished radial velocity data from STELLA \citep{granzer21}. 

The authors would also like to express their gratitude to Dr.~T.~Ayres for his guidance with accessing the ASTRAL archive. 

%\section*{Note}
%More details about the data, such as which spectral lines were used, can be found at \citet{bp}.


%\bibliographystyle{cas-model2-names} 
%\bibliographystyle{elsarticle-harv} 
%\bibliography{bibliography}


%\begin{thebibliography}{00}


%\newpage



\appendix

%\section{}

\begin{table*}[htbp]
	\caption{ Spectral lines used for the radial velocity analysis. The line lists were created based on \citet{carp}, \citet{brandt} and \citet{freeshell}. The exact wavelengths of all the lines were taken from NIST \citep{nist} and are given in $\rm \mathring{A}$. Ultraviolet wavelengths are given in vacuum, while the optical ones are given in air.}
	\label{table:lines} 
	\centering
	\scalebox{1}[1]{
	\begin{tabular}{lc|lc|lc|lc|lc}
\hline
\multicolumn{4}{c|}{Chromosphere and Stellar Wind} &  \multicolumn{2}{c|}{Ultraviolet photospheric region} &  \multicolumn{4}{c}{Optical region}\\ 
\cline{0-9}     
\multicolumn{1}{c}{Line}  & \multicolumn{1}{c}{Wavelength} & \multicolumn{1}{c}{Line}  & \multicolumn{1}{c|}{Wavelength} & \multicolumn{1}{c}{Line}  & \multicolumn{1}{c|}{Wavelength} & \multicolumn{1}{c}{Line}  & \multicolumn{1}{c}{Wavelength} & \multicolumn{1}{c}{Line}  & \multicolumn{1}{c}{Wavelength}\\
\hline

Fe II	&	2280.81	&	Fe II	&	2600.30	&	Fe II	&	2905.22	&	Ca II	&	3933.66	&	Ti II	&	4501.27	\\
Fe II	&	2328.27	&	Mn II	&	2606.61	&	Fe I	&	2913.01	&	Ca II	&	3968.47	&	Ti I	&	4512.73	\\
Fe II	&	2332.15	&	Fe II	&	2608.05	&	Fe I	&	2929.86	&	Ti I	&	4024.57	&	Ti I	&	4522.80	\\
Fe II	&	2333.66	&	Fe II	&	2612.04	&	Fe I	&	2937.76	&	Mn I	&	4030.76	&	Fe I	&	4920.50	\\
Fe II	&	2355.77	&	Fe II	&	2614.72	&	Fe I	&	2974.10	&	Mn I	&	4033.07	&	Ba II	&	4934.08	\\
Fe II	&	2369.48	&	Fe II	&	2618.58	&	Fe I	&	2982.31	&	Mn I	&	4034.49	&	Fe I	&	5079.74	\\
Fe II	&	2371.39	&	Fe II	&	2621.37	&	Fe I	&	2995.30	&	Fe I	&	4045.81	&	Mg I	&	5167.32	\\
Fe II	&	2374.58	&	Fe II	&	2622.63	&	Fe I	&	3008.16	&	Fe I	&	4063.59	&	Mg I	&	5183.60	\\
Fe II	&	2376.12	&	Fe II	&	2626.62	&	Fe I	&	3009.01	&	Fe I	&	4071.74	&	Fe I	&	5371.49	\\
Fe II	&	2380.20	&	Fe II	&	2629.25	&	Fe I	&	3014.37	&	Sr II	&	4077.71	&	Fe I 	&	5405.77	\\
Fe II	&	2385.27	&	Fe II	&	2715.36	&	Fe I	&	3018.51	&	Fe I	&	4132.06	&	Fe I	&	5429.70	\\
Fe II	&	2385.86	&	Fe II	&	2725.93	&	Fe I	&	3024.91	&	Fe I	&	4134.42	&	Ni I 	&	5476.91	\\
Fe II	&	2392.41	&	Fe II	&	2728.52	&	Fe I	&	3026.72	&	Fe I	&	4143.87	&	Na I	&	5889.95	\\
Fe II	&	2400.10	&	Fe II	&	2731.75	&	Fe I	&	3048.49	&	Fe I	&	4181.75	&	Na I	&	5895.92	\\
Fe II	&	2403.49	&	Fe II	&	2733.47	&	Ni I	&	3051.71	&	Fe I	&	4187.04	&	Fe I	&	6400.32	\\
Fe II	&	2450.62	&	Fe II	&	2737.97	&	Fe I	&	3059.98	&	V I	&	4190.73	&	Ti I	&	6413.10	\\
Fe II	&	2451.99	&	Fe II	&	2740.53	&	Ti II	&	3067.11	&	V I	&	4191.52	&	Fe I	&	6421.35	\\
Fe II	&	2485.10	&	Fe II	&	2744.19	&		&		&	Fe I	&	4199.10	&	Fe I	&	6430.85	\\
Fe II	&	2494.18	&	Fe II	&	2747.44	&		&		&	Ca I	&	4226.73	&	Ca I	&	6455.60	\\
Fe II	&	2506.15	&	Fe II	&	2747.91	&		&		&	V I	&	4234.00	&	Ca I	&	6462.57	\\
Fe II	&	2563.46	&	Fe II	&	2756.76	&		&		&	Cr I	&	4254.35	&	Fe I	&	6469.12	\\
Fe II	&	2564.37	&	Fe II	&	2760.37	&		&		&	V I	&	4259.31	&	Ca I	&	6471.66	\\
Fe II	&	2567.83	&	Fe II	&	2762.86	&		&		&	Fe I	&	4271.76	&		&		\\
Fe II	&	2578.85	&	Fe II	&	2769.95	&		&		&	Cr I	&	4274.81	&		&		\\
Fe II	&	2583.49	&	Mg II	&	2796.48	&		&		&	Cr I	&	4289.73	&		&		\\
Fe II	&	2586.80	&	Mg II	&	2803.67	&		&		&	Fe I 	&	4383.55	&		&		\\
Fe II	&	2592.50	&	Mg I	&	2853.15	&		&		&	Fe I	&	4466.55	&		&		\\
Mn II	&	2594.67	&	Fe II 	&	2881.81	&		&		&	Fe I	&	4482.17	&		&		\\
Fe II	&	2599.32	&		&		&		&		&	Fe I	&	4489.74	&		&		\\

	\hline
	\end{tabular}}
\end{table*}

\begin{table*}[htbp]
	\caption{Determined radial velocity $ v_{ \rm r} $ from all the spectra, used in the upper panel of Figure \ref{fig:vr_vrot}. All the UV radial velocities in the table were determined from the central subexposures. Omitted spectra - OEDQ04, OEDQ03 (subexposures 10-40), OEDQ02, OE1I02, ODXG01 (subexposure 10), ODXG04, ODXG03, ODXG05, OBKK71, O6LX03, O6LX02, as well as ZOYL01 (GHRS) and OBKK74 due to much larger apertures (0.2 $ \times $ 0.2$''$) }
	\label{table:v_r_results}
	\centering
	\scalebox{1}[1]{
	\begin{tabular}{llllll}
\textbf{Ultraviolet spectra}\\
\hline
\text{Source} & \text{MJD [d]} & \text{Dataset/Subexposure}  & \text{ $  v_{ \rm r, wind }  [\rm km \, s ^{-1} $]}  & \text{ $  v_{ \rm r, photo}  [\rm km \, s ^{-1} $]} & \text{  Aperture [$''$]} \\
\hline

%HST GHRS\footnotemark[2]  &	48889.12 & ZOYL01 &	14.49 $ \pm $ 0.79 & 18.46 $ \pm $ 1.02 & 0.25 $ \times $ 0.25  \\
HST ASTRAL &	50821.45  & O4DE03050 & 17.88 $ \pm $	0.48 & 20.92 $ \pm $	0.69 & 0.1 $ \times $  0.03  \\	
HST ASTRAL  &	50904.67 & O4DE05050 &	20.36 $ \pm $ 0.47 & 20.69 $ \pm $	0.73 & 0.1 $ \times $  0.03 \\	
HST ASTRAL  &	51077.18 & O4DE07050 &	21.06 $ \pm $	0.49 &	19.90 $ \pm $	0.75 & 0.1 $ \times $  0.03 \\	
HST ASTRAL  &	51265.31 & O4DE09050 &	18.73 $ \pm $	0.39 &	21.94 $ \pm $	0.69 & 0.1 $ \times $  0.03 \\	
HST ASTRAL  &	52562.57 & O6LX01020 &	19.27 $ \pm $	0.45 &	20.05 $ \pm $	0.94 & 0.2 $ \times $  0.06 \\	
%HST ASTRAL & o6lx02 &	52701.57 &	20.42 $ \pm $	0.41 &	\\
%HST ASTRAL & o6lx03 &	52752.54 &	20.28 $ \pm $	0.44 &	\\			
HST ASTRAL\footnotemark[2]  &	55600.34 & OBKK72010/72020 &	21.85 $ \pm $	0.47  & & 0.2 $ \times $  0.09 \\
%HST ASTRAL  &	55600.39 & OBKK72020 &	21.61 $ \pm $	0.74  & & 0.2 $ \times $  0.09 		\\	
%HST ASTRAL & obkk71 &	55606.12 &	20.21 $ \pm $	0.47 &				\\
HST ASTRAL\footnotemark[2]  &	55615.85 & OBKK73010/74020 &	21.84 $ \pm $	0.44 & & 0.2 $ \times $  0.09 	 			\\
%HST ASTRAL\footnotemark[2]  &	55617.91 & OBKK74010 &	20.38 $ \pm $	0.47 & 21.76 $ \pm $	0.67 &	0.2 $ \times $  0.20			\\
%HST ASTRAL  &	55617.96 & OBKK74020 &	21.72 $ \pm $	0.57 & &	0.2 $ \times $  0.09 			\\
HST ASTRAL  &	55652.38 & OBKK76020 &	21.45 $ \pm $	0.52 & & 0.2 $ \times $  0.09 				\\
HST STIS  &	58508.31  & ODXG01040 & 	19.33 $ \pm $	0.44 & 	21.64 $ \pm $	0.99 & 0.1 $ \times $  0.03				\\
HST STIS &	58547.57 & ODXG02040  &	18.68 $ \pm $	0.39 &	20.01 $ \pm $	0.78 &	 0.1 $ \times $  0.03			\\
%HST STIS  &	58573.06 & ODXG06 &	17.46 $ \pm $	0.46 &	0.1 $ \times $  0.03			\\
%HST STIS & odxg03 &	58591.34 &	17.57 $ \pm $	 0.59 &				\\
%HST STIS & odxg04 &	58714.76 &	17.30 $ \pm $	0.64 &				\\
HST STIS &	58744.33 & ODXG07040  &	17.67 $ \pm $	0.40 &	19.10 $ \pm $	1.10 &	0.1 $ \times $  0.03			\\
HST STIS  &	58762.81 & ODXG08040 &	16.85 $ \pm $	0.38 &	18.96 $ \pm $	1.24 &	0.1 $ \times $  0.03			\\

HST STIS  &	58815.64 & OE1I01040 &	18.21 $ \pm $	0.44 &	20.20 $ \pm $	1.12 &	0.1 $ \times $  0.03			\\
%HST STIS & oe1i02 &	58866.56 &	16.36 $ \pm $	0.59 &				\\

HST STIS  &	58882.38 & OE1I52040 &	15.75 $ \pm $ 0.37 &	16.67 $ \pm $ 0.83 &	0.1 $ \times $  0.03			\\

HST STIS  &	58904.82 & OE1I03040 &	15.82 $ \pm $	0.34 &	17.74 $ \pm $	0.72 &	0.1 $ \times $  0.03			\\

HST STIS  &	58941.04 & OE1I04040 &	14.50 $ \pm $	0.37  &	19.38 $ \pm $	0.75 &	0.1 $ \times $  0.03			\\

HST STIS  &	59092.77 & OEDQ01040 &	16.03 $ \pm $	0.36 &	19.91 $ \pm $ 0.65 &	0.1 $ \times $  0.03			\\
%HST STIS & oedq02 &	59137.85 &	17.88 $ \pm $	0.61 &				\\
HST STIS  &	59177.04 & OEDQ52040 &	16.46 $ \pm $   0.35 &	20.98 $ \pm $   0.84	 & 0.1 $ \times $  0.03				\\

%HST STIS & oedq03 &	59254.35 &	18.92 $ \pm $  0.60 	 &				\\
%HST STIS & oedq04 &	59326.89 &	17.73 $ \pm $	0.83 &				\\
HST STIS  &	59475.69 & OEDQ54040 &	19.13 $ \pm $   0.49 &	24.35 $ \pm $   1.18	 & 0.1 $ \times $  0.03 \\
 	\hline
%\midrule

 \\

\end{tabular}}
\scalebox{1}[1]{
	\begin{tabular}{llll}
 \textbf{Optical spectra} \\
\hline
\text{Source} &  \text{MJD [d]} & \text{ $ v_{ \rm r, OPT } $ [$\rm km \, s ^{-1} $]} & \text{  Aperture [$''$]}  \\
\hline
F/H	& 49868.96	& 19.58 $ \pm $	0.87 & 2.7 \\
F/H &	50455.68 &	18.37 $ \pm $	0.50 & 2.7 \\
ELODIE &	51204.90 &	22.50 $ \pm $	0.49 & 2 \\
UVES &	52530.42 &	21.54 $ \pm $	0.52 & 0.5 \\
ELODIE &	52873.14 &	22.82 $ \pm $	0.45 & 2 \\
ELODIE &	52899.16 &	22.56 $ \pm $	0.46 & 2 \\
ELODIE &	52953.09 &	20.42 $ \pm $	0.50 & 2 \\
ELODIE &	53048.90 &	19.33 $ \pm $	0.47 & 2 \\
ELODIE &	53121.81 &	20.56 $ \pm $	0.42 & 2 \\
ELODIE &	53244.16 &	22.75 $ \pm $	0.43 & 2 \\
ELODIE &	53404.90 &	18.84 $ \pm $	0.42 & 2 \\
SOPHIE &	54074.01 &	22.17 $ \pm $	0.49 & 3 \\
X{-}shooter &	55116.34 &	25.97 $ \pm $	0.67 & 0.5 \\
UVES & 	56570.38 &	20.18 $ \pm $	0.42 &  0.4\\
HARPS &	58891.07 &	19.83 $ \pm $	0.51 & 1		\\	

	\hline
     
	\end{tabular}}
	\\
	\footnotesize{$^2$ In these datasets the wavelength regions were split into 2378{-}2650 and 2621{-}2887, therefore results from these two regions were merged (when available). In the case of OBKK73010/74020 there was a 2 day gap between the observations. However considering the length of periods, this should not affect the results. The date of the latter observation is 55617.96.} \\
\end{table*}

%\footnotetext[2]{Unused in the analysis due to a different aperture.} 

\begin{table*}[htbp]
	\caption{Determined UV radial velocities across the disk of the Betelgeuse, used in the bottom panel of Figure \ref{fig:vr_vrot}. MJD of datasets correspond to the previous table. Position angle of the axis of rotation as determined by \citet{kervella} is 48$^{\circ}$ (all angles are given in $^{\circ}$ E of N). Position angles of the aperture given for each dataset. Subexposures correspond to different positions the ultraviolet surface of Betelgeuse, as discussed in Chapter \ref{chap:spec}. Some datasets also have 080 subexposures at distance of 100 mas from the centre. }
	\label{table:v_rot_results} 
	\centering
	\scalebox{1}[1]{
	\hskip-1.5cm
	\begin{tabular}{lc|c|c|c|c|c|c|c}
	
\multicolumn{9}{c}{Ultraviolet Photospheric Region}\\
\hline
\multirow{3}{*}{Dataset} & \multirow{3}{*}{ Position Angle [$^{\circ}$] }  &  \multicolumn{7}{c}{$ v_{ \rm r }  [\rm km \, s ^{-1}$]}\\ 
 \cline{3-9}  
& & \multicolumn{7}{c}{Subexposure}  \\
        \cline{3-9}     
& & \multicolumn{1}{c}{010}  & \multicolumn{1}{c}{020} & \multicolumn{1}{c|}{030}  & \multicolumn{1}{c}{050} & \multicolumn{1}{c}{060} & \multicolumn{1}{c}{070}& \multicolumn{1}{c}{080}\\
\hline

ODXG01 & 35.1 &  &  18.63 $ \pm $ 1.10	& 18.74 $ \pm $ 1.05 & 19.21 $ \pm $ 1.00 & 18.90 $ \pm $ 0.99 & 20.06 $ \pm $ 0.96 \\

ODXG02 & 40.5 &	22.13 $ \pm $ 	0.78	&	&	 &	&	&	18.78	$ \pm $  1.01  \\

ODXG06\footnotemark[3] & 52.5 &	19.40 $ \pm $ 	0.59	&		& &				& & \\
%odxg03	& 58591.27 &	18.56 $ \pm $ 	0.64	&		19.13 $ \pm $ 	0.59 &	17.63 $ \pm $ 	0.55 \\
%odxg04	& 58714.68 &	17.15 $ \pm $ 	0.76	&	&			17.53 $ \pm $ 	0.67 \\
ODXG07	 & -136.6 &	19.14 $ \pm $ 0.95	&	& & &	 &	19.13 $ \pm $ 	1.05 \\

ODXG08	 & -131.5 &	17.32 $ \pm $ 	0.86	&	&	 &	&	&	17.94	$ \pm $  1.11\\

OE1I01	 & -102.4 &	18.50 $ \pm $ 0.93	&	18.23 $ \pm $ 1.09 & 17.88 $ \pm $ 1.05 & 17.64 $ \pm $ 1.08 & 17.57 $ \pm $ 1.36 & 18.61 $ \pm $ 1.10 & 17.56 $ \pm $ 1.27 \\
%oe1i02	& 58866.56 &	16.36 $ \pm $ 	0.59	&					& \\
OE1I52 & 35.2  &	16.45 $ \pm $ 	0.87	&	&	 &	&	&	18.16	$ \pm $  0.81\\

OE1I03	& 38.0  &	17.35 $ \pm $ 	0.72	&	&	 &	&	&	18.68	$ \pm $  0.91 \\

OE1I04 & 50.5 &	18.57 $ \pm $ 	0.92	&	&	 &	&	&	17.89	$ \pm $  0.66\\

OEDQ01 & -141.4 &	19.56 $ \pm $ 0.87	&	18.13 $ \pm $ 0.95 & 17.75 $ \pm $ 0.91 & 18.61 $ \pm $ 1.02 & 18.34 $ \pm $ 1.11 & 19.07 $ \pm $ 0.77 & 18.17 $ \pm $ 1.02 \\

%oedq02	& 59137.76 &	17.45 $ \pm $ 	0.64	&		18.22 $ \pm $ 	0.64 &	19.03 $ \pm $ 	0.55 \\
OEDQ52   & -106.9 &    19.60 $ \pm $   0.84    & & & & &		20.13 $ \pm $   0.85 	 	\\

OEDQ03\footnotemark[3]   & 32.2 & 	&	&	 &		& &	21.97	$ \pm $  0.82\\
%oedq04	& 59326.84 &	17.73 $ \pm $ 	0.83	&					\\
%OEDQ54 & -136.4  & 	&	&	 &		& &	\\
	\hline
	\end{tabular}}
	\footnotesize{$^3$ Unused in bottom panel of Figure \ref{fig:vr_vrot} due to a missing subexposure from the centre of the star.}\\
	
\end{table*}

\end{document}